\documentclass[fleqn,10pt]{wlscirep}

\usepackage{amsmath,graphicx,subcaption,graphics}
\usepackage[export]{adjustbox}
\newcommand\numberthis{\addtocounter{equation}{1}\tag{\theequation}}

\newcommand{\avg}[1]{\left\langle #1\right\rangle}

\newcommand{\T}[1]{\textrm{#1}}

\title{Competitive percolation strategies for network recovery}

\author[1,*]{Andrew M. Smith}
\author[1]{M{\'a}rton P{\'o}sfai}
\author[1]{Martin Rohden}
\author[2]{Andr{\'e}s D. Gonz{\'a}lez}
\author[3]{Leonardo Due{\~n}as-Osorio}
\author[1,4,5]{Raissa M. D'Souza}
\affil[1]{Department of Computer Science and Complexity Sciences Center, University of California, Davis, CA, USA, 95616}
\affil[2]{School of Industrial and Systems Engineering, University of Oklahoma, Norman, OK, USA, 73019}
\affil[3]{Department of Civil \& Environmental Engineering, Rice University, Houston, TX, USA, 77005}
\affil[4]{Department of Mechanical and Aerospace Engineering, University of California, Davis, California, USA, 95616}
\affil[5]{Santa Fe Institute, Santa Fe, New Mexico, USA, 87501}

\affil[*]{amsmi@ucdavis.edu}

\begin{abstract}
Restoring operation of critical infrastructure systems after catastrophic events is an important issue, inspiring work in multiple fields, including network science, civil engineering, and operations research. We consider the problem of finding the optimal order of repairing elements in power grids and similar infrastructure. Most existing methods either only consider system network structure, potentially ignoring important features, or incorporate component level details leading to complex optimization problems with limited scalability. We aim to narrow the gap between the two approaches. Analyzing realistic recovery strategies, we identify over- and undersupply penalties of commodities as primary contributions to reconstruction cost, and we demonstrate traditional network science methods, which maximize the largest connected component, are cost inefficient. We propose a novel competitive percolation recovery model accounting for node demand and supply, and network structure. Our model well approximates realistic recovery strategies, suppressing growth of the largest connected component through a process analogous to explosive percolation. Using synthetic power grids, we investigate the effect of network characteristics on recovery process efficiency. We learn that high structural redundancy enables reduced total cost and faster recovery, however, requires more information at each recovery step. We also confirm that decentralized supply in networks generally benefits recovery efforts.
\end{abstract}
\begin{document}

\flushbottom
\maketitle

\thispagestyle{empty}

\section*{Introduction}
Resilience of complex networks is one of the most studied topics of network science, with an expanding literature on spreading of failures, mitigation of damage, and recovery processes \cite{Albert2000a,Cohen2000,Motter2002,Li2014,Zhao2016,Zhong2018}. The level of functionality of a network is typically quantified by its connectedness, e.g., size of the largest component~\cite{Albert2000a}, average path length~\cite{Latora2001a,Ash2007a}, or various centrality metrics~\cite{Holme2002a}. Such simple topology-based metrics ensure mathematical tractability and allow us to analyze and compare networks that can be very different in nature, providing general insights into the organization of complex systems. However, such a perspective necessarily ignores important system-specific details. For example, abstracted topological models of infrastructure networks recovering from damage or catastrophic failure aim to rapidly restore the largest component~\cite{Majdandzic2014a,Hu2016a,Shang2016a,Muro2016a}. But, extensive connectivity is not a necessary condition to guarantee that all supply and demand can be met. For instance, consumers of a power grid can be served if they are connected to at least one power source and that source satisfies operational constraints~\cite{Quattrociocchi2014}. The concept of ``islanding'', a technique of intentionally partitioning the network to avoid cascading failures, is actually a practical strategy used to improve security and resilience during restoration efforts in power grids~\cite{Panteli2016a,Mureddu2016a,NRC2012a}. Indeed, after the 2010 earthquake in Chile the recovery process first created five islands, which were only connected to each other in the final steps of reconstruction~\cite{Rudnick2011a}.

The restoration of critical infrastructure operation after a catastrophic event, such as a hurricane or earthquake, is a problem of great practical importance and is the focus of a significant body of work in civil and industrial engineering disciplines. The goal of engineering based models of recovery is to provide system-specific predictions and actionable recommendations. This is achieved by incorporating component level details and realistic transmission dynamics into the models, often in the form of generalized formulations of the network design problems (NDPs), which satisfy network flows. In this context, objective functions of such NDP-based models aim to minimize the construction and/or operational costs of recovering edges and nodes in a utility network. Basic forms of the NDP have been well-studied \cite{Johnson1978a,Balakrishnan1989a}, and have recently been combined with scheduling and resource allocation problems to model the entire restoration process~\cite{Nurre2012a,Gonzalez2016a,Gonzalez2016b}. While these models provide a principled manner to obtain optimal, centralized recovery strategies, their complexity (at least NP-complete~\cite{Johnson1978a}) renders computation not scalable, and interpretation restricted in scope to small instances. More efficient approximate solutions for NDPs have been found using optimization meta-heuristics such as a hybrid ant system~\cite{Poorzahedy2007a} and gradient descent~\cite{Gallo2010a} methods. Such algorithms are generally applicable to global search problems and were designed to reduce computational complexity by not guaranteeing optimality; therefore, provide limited insight into the mechanism of network formation during recovery. We will analyze the output of an NDP algorithm, and leveraging on these observations, we will develop a percolation-based model for network recovery with the goal of uncovering important principles of network formation and recoverability.

\textit{Percolation processes}, 
often used for studying properties of stochastic network formation, have recently been applied to network recovery~\cite{Li2015}. In the kinetic formulation of random percolation~\cite{krapivsky2010}, we start with $N$ unconnected nodes and consider a discrete time process. At each timestep, an edge is selected from the set of all possible edges at random, and added to the network. Initially the largest connected component (LCC) is sublinear in $N$; above a critical edge density it spans a finite fraction of the network and the LCC is referred to as the giant component. Controlling location of the critical point is of great interest in many systems -- for instance, suppressing the formation of the giant component may reduce the likelihood of virus spreading in social contact networks. This can be achieved by selecting $M > 1$ candidate edges at each timestep, and adding the edge that optimizes some criteria. The general class of models that results from this choice is referred to as competitive percolation or an \textit{Achlioptas process}~\cite{Achlioptas2009,DSouza2015a}. While simple, Achlioptas processes often have the benefit of being scalable, numerically analyzable, and provide a parameter, $M$, for tuning how close the formation process is to matching the desired criteria. Note that when $M$ is equal to the number of possible edges,  we always add the 
edge that is optimal with respect to the selection criteria. Previous percolation-based recovery models typically measure solution quality by how quickly the LCC grows~\cite{Hu2016a,Shang2016a}, or assume nodes which are not connected to the LCC to be nonfunctional~\cite{Majdandzic2014a,Muro2016a}. However, empirical studies of recovery scenarios suggest that these assumptions do not apply to infrastructure networks after catastrophic scenarios~\cite{Rudnick2011a} . 

In this work, we aim to narrow the gap between topology-based recovery approaches and computationally difficult optimization approaches by incorporating features which mirror infrastructure restoration processes. We start by applying a generalized version of a well-studied NDP recovery algorithm~\cite{Gonzalez2016a,Gonzalez2016b,Gonzalez2017a} to a small case study, and  we identify that the satisfaction of demand is a key driving force in the initial periods of recovery, outranking operational efficiency and direct repair costs of network elements in importance. Motivated by this finding, we define a simple, competitive percolation-based model of recovery that aims to maximize the satisfaction of consumer demand in a greedy manner. We show that component size anti-correlates with the likelihood of further growth; therefore, leading to islanding and the suppression of the emergence of large-scale connectivity, similar to explosive percolation transitions~\cite{DSouza2015a} and in contrast with traditional recovery models. We apply our recovery algorithm to synthetic power grids to systematically investigate how realistic structural features of the network affect the efficiency of the recovery process. We learn that high structural redundancy (related to the existence of multiple paths between nodes) allows for reduced total cost and faster recovery time; however, to benefit from that redundancy, an increasing amount of information needs to be considered at each step of reconstruction. We also study the role of the ratio of suppliers and consumers and find that decentralized supply generally benefits recovery efforts, unless the fraction of suppliers becomes unrealistically high. Our model deepens our understanding of network formation during recovery and of the relationship between network structure and recoverability. We anticipate that our work can lead to efficient approximations of the NDP algorithm by leveraging the important mechanisms uncovered by our competitive percolation model.

\section*{Model}

\subsection*{Problem statement and the optimal recovery model}

We are interested in the problem of restoring the operation of a critical infrastructure system after sustaining large-scale damage. The infrastructure network is represented by a graph $\mathcal{G}=(\mathcal{N},\mathcal{E})$, where $\mathcal{N}$ is the set of $N$ nodes corresponding to substations and $\mathcal{E}$ is a set of $E$ edges corresponding to transmission lines, e.g., power lines, water or gas pipes. We introduce the parameter $d_i$ representing the commodity demand of node $i$: if $d_i<0$, node $i$ is a net consumer; if $d_i>0$, node $i$ is a net supplier. We normalize $d_i$ such that the total consumption (or production) sums to unity, i.e., $\frac{1}{2}\sum_{i=1}^N \left| d_i \right|=1$. Following a catastrophic event, a subset of the network $\mathcal{G}^{\prime}=(\mathcal{N}^{\prime},\mathcal{E}^{\prime})$ becomes damaged. We study a discrete time reconstruction process: in each timestep we fix one damaged component, and the process ends once the entire network is functional. Our goal is to identify a sequence in which to repair the elements such that the total cost of recovery is minimized. In this manuscript, we focus on the fundamental case where all links are damaged but nodes remain functional, i.e., $\mathcal{E}^{\prime}=\mathcal{E}$ and $\mathcal{N}^{\prime}=\emptyset$.

Optimization frameworks are often used in order to explore the space of possible repair sequences and identify best solutions. We implement the time-dependent network design problem (td-NDP) algorithm~\cite{Gonzalez2016a,Gonzalez2016b,Gonzalez2017a}, which is a well known example of an optimization  algorithm for network recovery developed by the civil engineering community. Out of the recovery processes we examine in this paper, td-NDP is the most realistic, and therefore the most computationally complex. It is formulated as a mixed integer problem which optimizes a cost that includes reconstruction costs of network components, operational costs, and penalties incurred for unsatisfied demand, while taking constraints on flows of commodities into account. In general mixed integer program are known to be NP-hard except in special cases. For the td-NDP this means problems become exponentially harder as the size of the network to be reconstructed increases; therefore, it is common practice to break up the recovery into time windows of length $T$, and find the locally optimal solution in each window. A formal definition of td-NDP is provided in the Methods section.

To uncover the key driving factors and properties of the recovery process, we apply the td-NDP to a representative example, the power grid of Shelby County, Tennessee, which consists of $9$ suppliers and $37$ consumers (with $14$ junction nodes where $d_i = 0$), connected by $M=76$ transmission lines. Network topology and necessary parameters where obtained from Refs.~\cite{Gonzalez2016a,Gonzalez2016b}. Figure \ref{fig:breakdown}a shows the total repair cost as a function of time as we perform the td-NDP with $T = 5$ on a network that was initially completely destroyed, with cost broken down by the type of expense. We see that deficit cost (i.e., the penalty accrued for unsatisfied demand) is overwhelmingly dominant and exponentially decreasing throughout the initial stages of recovery. Investigating how this impacts the growth of the components, Figure \ref{fig:sizeVsSurplus}a shows the commodity deficit or supply of each connected component throughout the recovery process, with circle sizes representing the component size, and colors representing if a component is over- (blue) or undersupplied (red). We see that the td-NDP process results in many small components initially, with relatively small surplus/deficit, and only towards the end of the process all components are joined. This is consistent with islanding techniques discussed in engineering practice. Our goal is to develop a simple and computationally efficient model of the recovery process that captures these key features.

\subsection*{Competitive percolation optimizing for LCC growth}
Previous topology-based recovery processes prioritize the rapid growth of the Largest Connected Component (LCC)~\cite{Hu2016a,Shang2016a,Majdandzic2014a,Muro2016a}. Models vary in the details, such as the type of failure (random, localized, catastrophic) and additional secondary objectives (such as prioritizing nodes based on population), but the metric for the quality of the solution is directly related to how quickly the LCC grows. 

As a representative example of topology driven recovery strategies, we implement an Achiloptas process using a selection rule that maximizes the sum of the resulting component, which we refer to as \textit{LCC percolation}. In this process, we randomly select $M > 0$ candidate edges out of the set of damaged edges at each discrete timestep. We then examine the impact that each individual edge would have and select the edge that, when added, creates the largest connected component. More specifically, let $s_i$ denote the size of the component to which node $i$ belongs. If nodes $i$ and $j$ belong to separate components, repairing edge $(i,j)$ creates a component with size $S_{ij}=s_i+s_j$; if they belong to the same component, the size of the component does not change and we set $S_{ij}=0$. Out of the $M$ candidate edges, we repair the one that maximizes $S_{ij}$; if multiple candidate links have the same maximal $S_{ij}$, we  select one of them uniformly at random. If $M=1$, the process is equivalent to traditional percolation. If $M=E$, the process is largely deterministic, we always repair an edge that is optimal with respect to the selection criteria.

We now apply LCC percolation as a model for the recovery of the Shelby County power grid and compare the results to the benchmark td-NDP process. Figure \ref{fig:sizeVsSurplus}b shows that if we use growth of the LCC  as our objective, the LCC, represented by the largest circle, grows rapidly throughout the process as expected. However, the deficit/surplus of this component fluctuates greatly. As indicated by the magnitude of total commodity deficit $D$, i.e., the total unsatisfied demand in the network, in Figure \ref{fig:breakdown}b, such a recovery algorithm is costly and leaves large portions of the grid without power until the final steps of the recovery algorithm. To conclude, we find that a recovery process based on quick growth of the LCC is neither cost efficient nor effective for satisfying consumer's demand as quickly as possible. As a result, we do not consider this algorithm for further studies.

\subsection*{Competitive percolation optimizing for demand satisfaction}

We have shown that the key driving factor in recovery processes is reduction of the total commodity deficit, i.e., the total unsatisfied demand, and that optimizing LCC fails to capture this. To capture the essence of real recovery strategies, we propose a competitive percolation process which we refer to as {\it recovery percolation} that, instead of optimizing for LCC growth, aims to directly reduce the unsatisfied demand. In addition to network topology, this recovery process also takes into account the net demand or production of the individual nodes.

We define $D_i$ as the commodity deficit of the connected component to which node $i$ belongs. We assume that capacity constraints of the transmission lines are sufficient and thus do not limit the flow of commodities during the recovery process, a common practice in infrastructure recovery literature~\cite{Quattrociocchi2014,Gonzalez2016a,Gonzalez2016b}. Therefore the commodity deficit of a component is the sum of demand or supply of individual nodes belonging to the component, i.e., $D_i=\sum_{j\in \mathcal{C}_i}d_j$, where $\mathcal{C}_i$ is the set of nodes belonging to the component containing node $i$.

We use commodity deficit of the components as a selection criteria for the competitive percolation model to account for the goal of balancing supply and demand. Similar to LCC percolation discussed above, we randomly select $M > 0$ damaged candidate edges, from which one is chosen to be repaired and added to the network at each timestep as follows.  We first consider how much demand would be met by adding each of the $M$ edges individually to the network. More specifically, if nodes $i$ and $j$ belong to components such that $D_iD_j<0$, then repairing edge $(i,j)$ reduces the total commodity deficit by $\Delta D = \min(\lvert D_i\rvert, \lvert D_j\rvert)$; if $D_iD_j\geq 0$, then there is no commodity deficit reduction, i.e., $\Delta D= 0$. Out of the $M$ candidate edges, we repair the one that maximizes $\Delta D$; if multiple candidates have the same maximal $\Delta D$, we repair one of them chosen uniformly at random. As $M$ shrinks and approaches $1$, the process becomes more stochastic; while if $M=E$ the process is largely deterministic, as we always select an edge that is optimal with respect to the selection criteria.

\begin{figure}[ht!]
{\centering
	\includegraphics[width=1.0\textwidth]{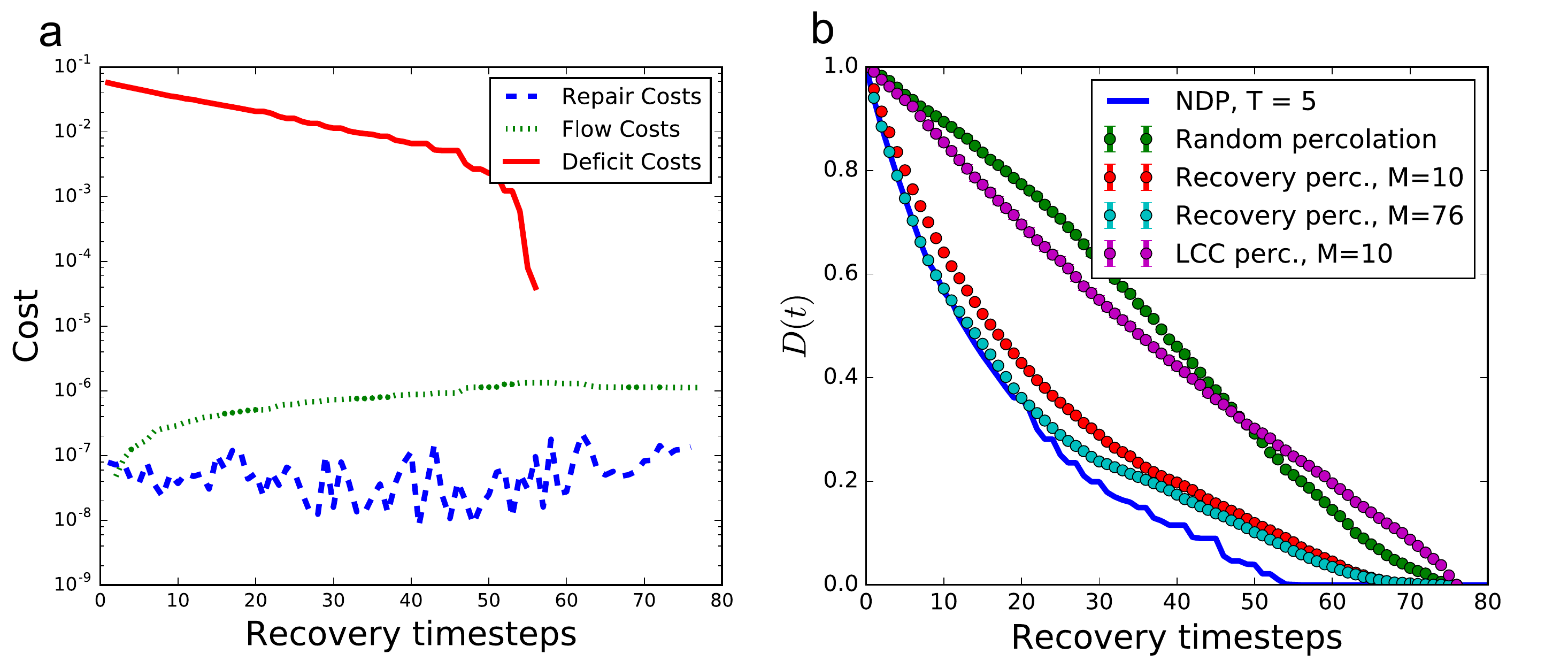}}
	\caption{Recovery of the Shelby County power network. (a)~Breakdown of normalized recovery costs for the locally optimal td-NDP solution. Surplus/deficit costs are the overwhelmingly dominant factor in early stages of recovery. (b)~Total commodity deficit during the recovery of the Shelby County power network, averaged over $100$ independent realizations for percolation results; recovery percolation model closely approximates the td-NDP baseline surplus/deficit cost curve  (especially as $M$ increases), while LCC percolation deviates.}
\label{fig:breakdown}
\end{figure}

\begin{figure}
\centering
	\includegraphics[width=1.0\textwidth]{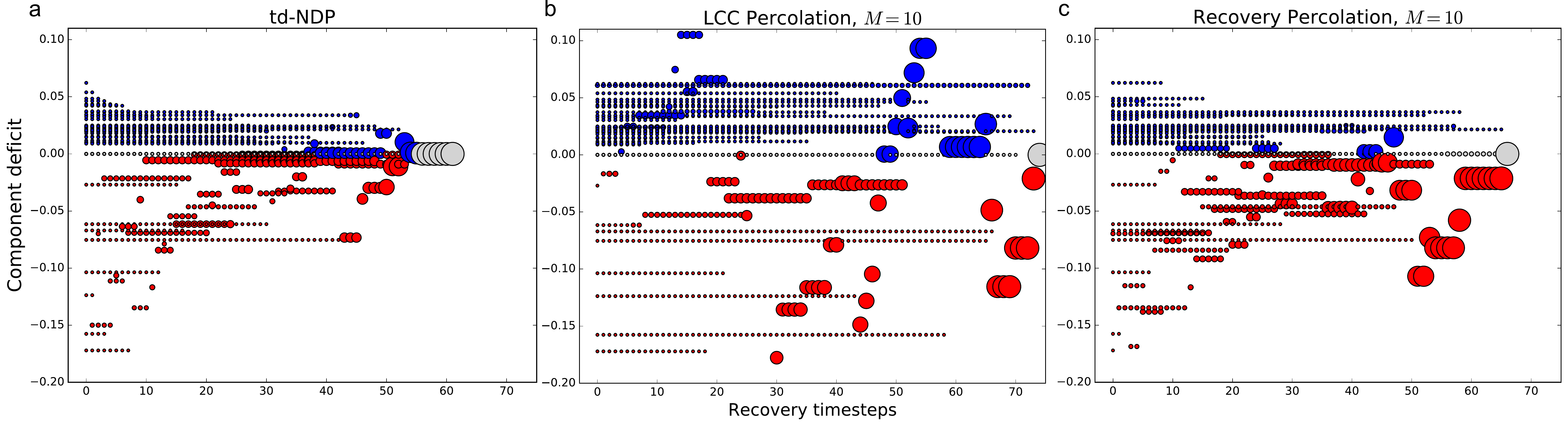}
	\caption{Commodity deficit of each connected component over time throughout various recovery processes on the Shelby County dataset: (a) td-NDP, (b) LCC percolation, and (c) recovery percolation. Each point represents a component, where the size of the point indicates the relative size of the component. Blue circles (above $0$ on the y-axis) indicate excess supply, while red circles (below $0$ on the y-axis) indicate unmet demand. The final point on the x-axis indicates the first timestep where all nodes belong to the LCC. The optimization process shown in (a)~keeps size and commodity deficit low throughout the recovery process, consistent with islanding practices. Standard percolation, (b), is vastly different, showing varying sizes and commodity deficits throughout the process, resulting in higher recovery costs. Our competitive recovery percolation method, (c), shows a signature closer to (a) and is much more computationally efficient, making a locally optimal choice from a sample size of only $M=10$ at each timestep.}
	\label{fig:sizeVsSurplus}
\end{figure}

Figure \ref{fig:breakdown}b shows that for the Shelby County power grid the total commodity deficit during the recovery percolation for $M=E=76$ well approximates the td-NDP, especially at the beginning of the recovery process when costs are much higher. We also see that even when $M=10$, corresponding to only $\sim 13\%$ of the total edges considered at each timestep, the approximation remains very effective. Figure \ref{fig:sizeVsSurplus}c shows similar dynamical behavior in recovery percolation as in the td-NDP solution (cf.~Figure \ref{fig:sizeVsSurplus}a): larger components delay formation, and tend to have smaller commodity deficit.

\section*{Results}
In the following, we apply the recovery percolation model to various network topologies to identify important mechanisms driving network formation and to understand how network structure affects the efficiency of recovery efforts. For each synthetically generated network, the demand distribution is chosen to approximate the demand observed in real power grids (details are provided in the Methods section).

\subsection*{Recovery percolation on complete networks}
We have shown that recovery percolation follows our benchmark td-NDP solution closely on a real-world topology. We also observed that the growth of connected components is suppressed via recovery percolation as compared to LCC percolation. To understand this behavior, we study large systems with $N=10^4$ nodes and we allow potential edges to exist between any node pair, removing underlying topology constraints. Note that the td-NDP process is intractable for networks of this size.

\begin{figure}
\centering
\includegraphics[width=1.0\textwidth]{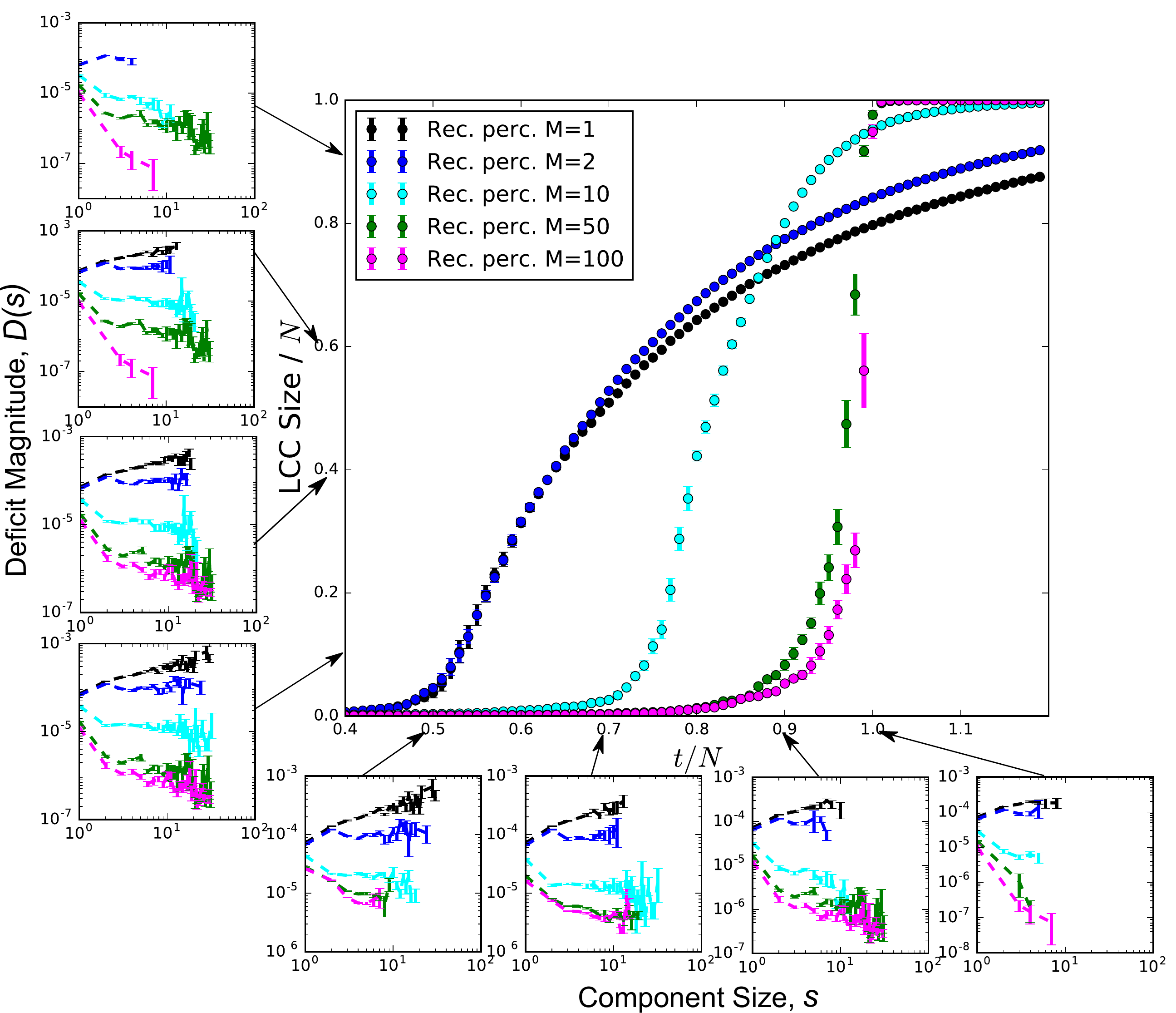}
\caption{Recovery percolation on the complete graph, showing growth of the LCC with increasing edge density (center figure). We see that as $M$ is increased, the transition point is delayed. Small plots show the average undersupply of connected components as a function of component size. Left column plots distributions for various $M$ values when the LCC is of a certain size; bottom row plots shows the same for fixed time. Rare events (component sizes that occur $< 1\%$ of the time) are not plotted to eliminate noise. We note that as $M$ is increased, the slopes trend towards being more negative, indicating that larger components have less deficit magnitude throughout the recovery process when more recovery choices are presented to the recovery percolation process. All data points are the average of 10 independent realizations with $N=10^4$ nodes, error bars indicate the standard error of the mean.}
\label{fig:perc_behaviors}
\end{figure}

Figure \ref{fig:perc_behaviors} (main figure) shows the growth of the LCC for a range of $M$ values. For $M=1$, 
the model reduces to random percolation which has a second-order phase transition at $t/N=0.5$, and above this critical point the LCC becomes proportional to $N$. As we increase $M$, the apparent transition point shifts to higher values and approaches $1$, indicating that the appearance of large-scale connectivity is suppressed; however, once the transition point is reached, the growth of the LCC becomes increasingly abrupt. This observation is analogous to explosive percolation, where links are chosen to be constructed explicitly to delay component growth~\cite{DSouza2015a}. In contrast, in recovery percolation it is an indirect consequence of the restoration strategy. 

To understand the underlying mechanism of component growth, we plot the average component sizes and their corresponding average undersupply at various points during the reconstruction process in the bottom row of plots in Figure \ref{fig:perc_behaviors}. Note that average oversupply behaves in a similar manner, but is omitted for clarity. The left column of plots in Figure \ref{fig:perc_behaviors} shows the same quantities such that the size of the LCC is fixed. The main trend we observe at any given point in time is that for large enough $M$ there is negative correlation between component size and undersupply, and this correlation becomes stronger as $M$ increases. This means that as components grow their commodity deficit is reduced and therefore the likelihood of further growth is also reduced, ultimately suppressing the appearance of large scale connectivity. 

The observed two features also describe islanding, an intentional behavior in resilience planning and recovery in real-world power grids. This islanding behavior is already observed in early stages of the restoration process, becoming more apparent as $t$ approaches the transition point.

\subsection*{Recovery percolation on synthetic power grids}
\label{sec:power_net_results}
So far we investigated the recovery process on an underlying graph without topological constraints. We also wish to analyze more realistic networks and turn our attention to synthetic power grids~\cite{wang2010,Schultz2014a}. This allows us to systematically investigate how typical structural features of power grids affect the efficiency of the recovery process.

Power grids are spatially embedded networks, and physical constraints limit the maximum number of connections a node can have; their degree distributions, therefore, have an exponential tail, in contrast to many complex networks that display high levels of degree heterogeneity. Power grids typically have average degree $\avg k$ between $2.5$ and $3$~\cite{Schultz2014a,Li2014b}. An important requirement of power grids is structural redundancy, meaning that the failure of a single link cannot cause the network to fall into disconnected components. A network without redundancy has tree structure, has average degree 2 and all node pairs are connected through a unique path. Any additional link creates loops and improves redundancy. Structural redundancy can be characterized locally by counting short range loops. For example, power grids have a high clustering coefficient $c$, typically ranging between $0.05$ and $0.1$~\cite{Schultz2014a}. The algebraic connectivity, denoted by $\lambda_2$, is the second smallest eigenvalue of the network's Laplacian matrix and captures a measure of global redundancy: it is related to the number of links that have to be removed in order to break the network into two similarly sized components, with high value corresponding to high redundancy. The exact value of $\lambda_2$ depends on system size, where for a given number of nodes $\lambda_2$ is minimal for tree structure, and monotonically increases as further links are added~\cite{Miegham2010a}.

To generate networks that exhibit these features, we use a simplified version of a practical model developed by Schultz et al.~\cite{Schultz2014a}. The model generates spatially embedded networks mimicking the growth of real-world power grids. The process is initiated by randomly placing $N_0$ nodes on the unit square and connecting them with their minimum spanning tree. To increase redundancy, $qN_0$ ($0\leq q\leq 1$) number of links are added one-by-one, such that each link is selected to minimize the redundancy-cost trade-off function
\begin{equation}
f(i,j)=\frac{(d_\T{net}(i,j)+1)^r}{d_\T{euc}(i,j)},
\end{equation}
where $i$ and $j$ are two nodes not connected directly, $d_\T{net}(i,j)$ is their shortest path distance in the network, and $d_\T{euc}(i,j)$ is their Euclidean distance. The $r\geq 0$ parameter controls the trade-off between creating long loops to improve redundancy and the cost of building power lines. After the initialization, we add $N-N_0$ nodes through a growth process. In each time step a new node is added: with probability $1-s$ the node is placed in a random position and connects to the nearest node; with probability $s$ a randomly selected link $(i,j)$ is split and a new node is placed halfway between nodes $i$ and $j$ and is connected to both of them. To increase redundancy, in each time step an additional link is added with probability $q$ connecting a randomly selected node $i$ to node $j$, such that $f(i,j)$ is minimized. Finally, a fraction of nodes $p_\T s$ are randomly selected to be suppliers, the rest are assigned to be consumers. 

\begin{figure}[t!]
	{\centering
	\includegraphics[width=1.0\textwidth]{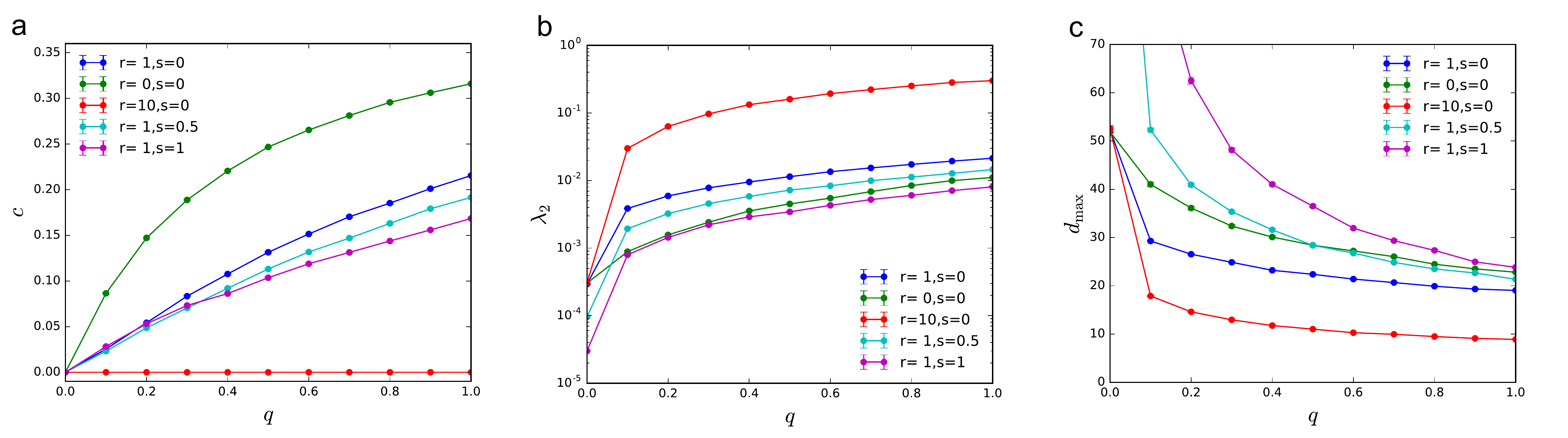}
	}	
	\caption{
	 Network structure of the synthetic power grids. (a-c)~Increasing $q$ increases both local and global redundancy, i.e., clustering $c$ and algebraic connectivity $\lambda_2$; while decreases the diameter $d_\T{max}$ of the network. High $r$ (solid red) favors long distance shortcuts, increasing $\lambda_2$ and decreasing $c$; while small $r$ (green) has the opposite effect. Increased line splitting $s$ (light blue and purple) increases $d_\T{max}$, especially for low $q$; while reducing both $c$ and $\lambda_2$. Each data point is an average of $100$ independent network realizations with $N=1000$ nodes, and the errorbars indicate the standard error of the mean.}
	\label{fig:properties}
\end{figure}

Changing parameters $q$, $r$, and $s$ allows us to systematically explore how these parameters impact the structure of these model power grids~(Fig.~\ref{fig:properties}): $q$ controls the average degree $\avg k = 2(1+q)$ and adds redundancy to the network; $r$ controls how loops are formed, where small $r$ favors short distance connections leading to high $c$ and low $\lambda_2$, while large $r$ favors long loops leading to low $c$ and high $\lambda_2$; and $s$ increases typical distances in the network, lowering both $c$ and $\lambda_2$.

\subsubsection*{Comparing recovery percolation and td-NDP}

\begin{figure}[t]
	{\centering
	\includegraphics[width=1.0\textwidth]{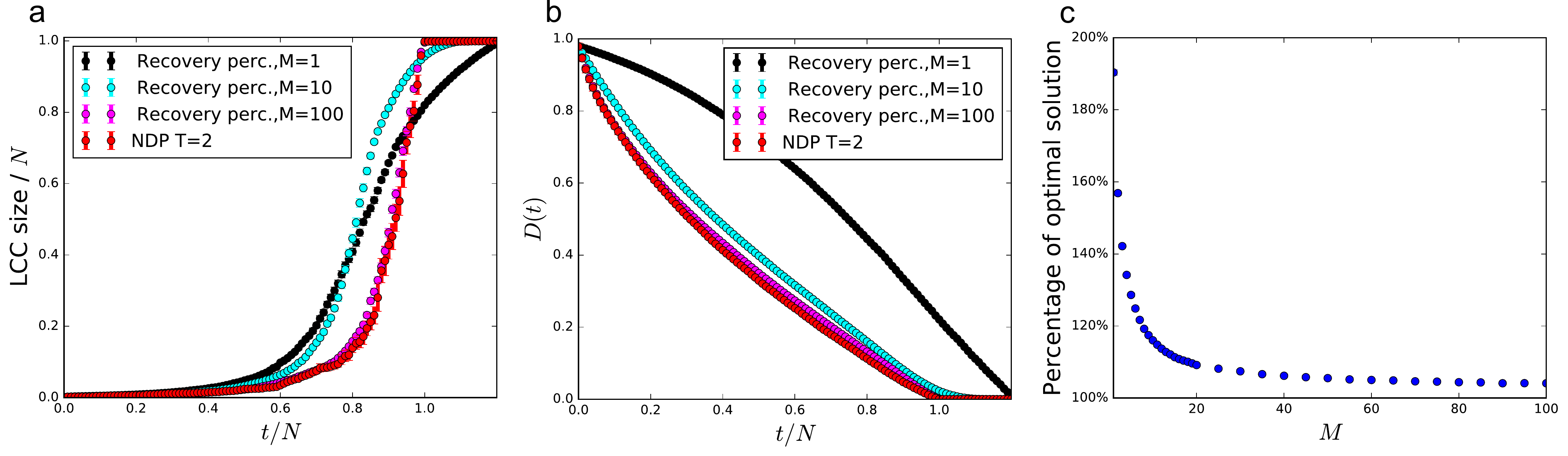}}
	\caption{Recovery percolation and td-NDP on synthetic power grids. (a)~Size of the LCC as a function of recovery time and (b)~total  commodity deficit during the restoration. High $M$ values well capture the behavior of the td-NDP process. (c)~How well the total cost during recovery percolation approximates the cost of the td-NDP solution as a function of $M$. All data points are the average of 10 independent realizations with $N=1000$ nodes, error bars indicate the standard error of the mean and are typically smaller than the markers.}
	\label{fig:power_results}
\end{figure}

We first consider a set of parameters that yield typical topologies and compare the performance of recovery percolation to that of the locally optimal td-NDP recovery. We choose the parameters to create networks similar to the Western US grid following the specifications of Ref.~\cite{Schultz2014a} ($N=10^3$, $q=0.33$, $r=1$, and $s=0$). For the td-NDP analysis, we reduce the time window from $T=5$, as used in the Shelby County model, down to $T=2$ for tractability reasons since our synthetic networks are much larger (increasing $T$ causes an exponential increase in complexity). While this will result in a suboptimal solution, it still considers future timesteps, a dimension not present in percolation models. Figure \ref{fig:power_results}a shows the growth of the LCC  for the td-NDP process and recovery percolation varying $M$ from $1$ to $100$. For recovery percolation we observe similar behavior to that seen for studies on the complete networks: as $M$ is increased the growth of the LCC is suppressed, and the formation of large-scale connectivity is delayed, and when it forms it grows more rapidly. For large $M$, the recovery percolation closely resembles the td-NDP recovery in terms of LCC formation. 

As the dominant cost factor in recovery of infrastructure networks is the total commodity deficit $D(t)$, this is the most important metric in network recoverability, beyond the size of the LCC. Figure \ref{fig:power_results}b shows $D(t)$ reduction throughout the recovery process. As $M$ increases, we see a closer fit with td-NDP, especially in the more expensive early stages of recovery. Surprisingly, $M=10$ approximates total commodity deficit quite well, which is significantly less computationally intensive than td-NDP or the deterministic version of recovery percolation ($M=E$).

To better understand how the choice of $M$ affects the quality of recovery percolation, we calculate the total cost $C_M$, defined here as the area under the curve $D(t)$ over time (i.e., $C_M=\sum_t D(t)$) as a function of $M$. Figure~\ref{fig:power_results}c shows that $C_M$ rapidly approaches $C_\infty$, its value at $M=E$. For this particular case, we only need to consider $M=20$, that is less than $2\%$ of edges, at each timestep to get within $10\%$ of the optimal cost.

It is worth highlighting that recovery percolation captures the essential properties of the td-NDP process for $T=2$, despite the fact that recovery percolation only considers commodity deficit, while td-NDP takes into account such details as heterogeneous repair costs of individual power lines, operational costs, performs network flows, and selects optimal recovery actions considering two timesteps.

\subsubsection*{Effect of network structure on recovery percolation}

Recovery percolation together with the synthetic power grid networks provide a stylized model to extract key network features that impact the efficiency of the restoration process. For this we systematically investigate how typical structural features affect the following quantities:
\begin{enumerate}
\item Total optimal cost of recovery $C_\infty$, which is the minimum cost obtainable with recovery percolation ($M=E$).
\item Time to recovery $t_{90}$ , the number of timesteps needed to reduce total commodity deficit by 90\% percent.
\item Characteristic $M^*$, which captures the approximability of the process. It is defined as the smallest $M$ value for which $C_M\leq 1.2 C_\infty$.
\end{enumerate}

\begin{figure}[t!]
	{\centering
	\includegraphics[width=1.0\textwidth]{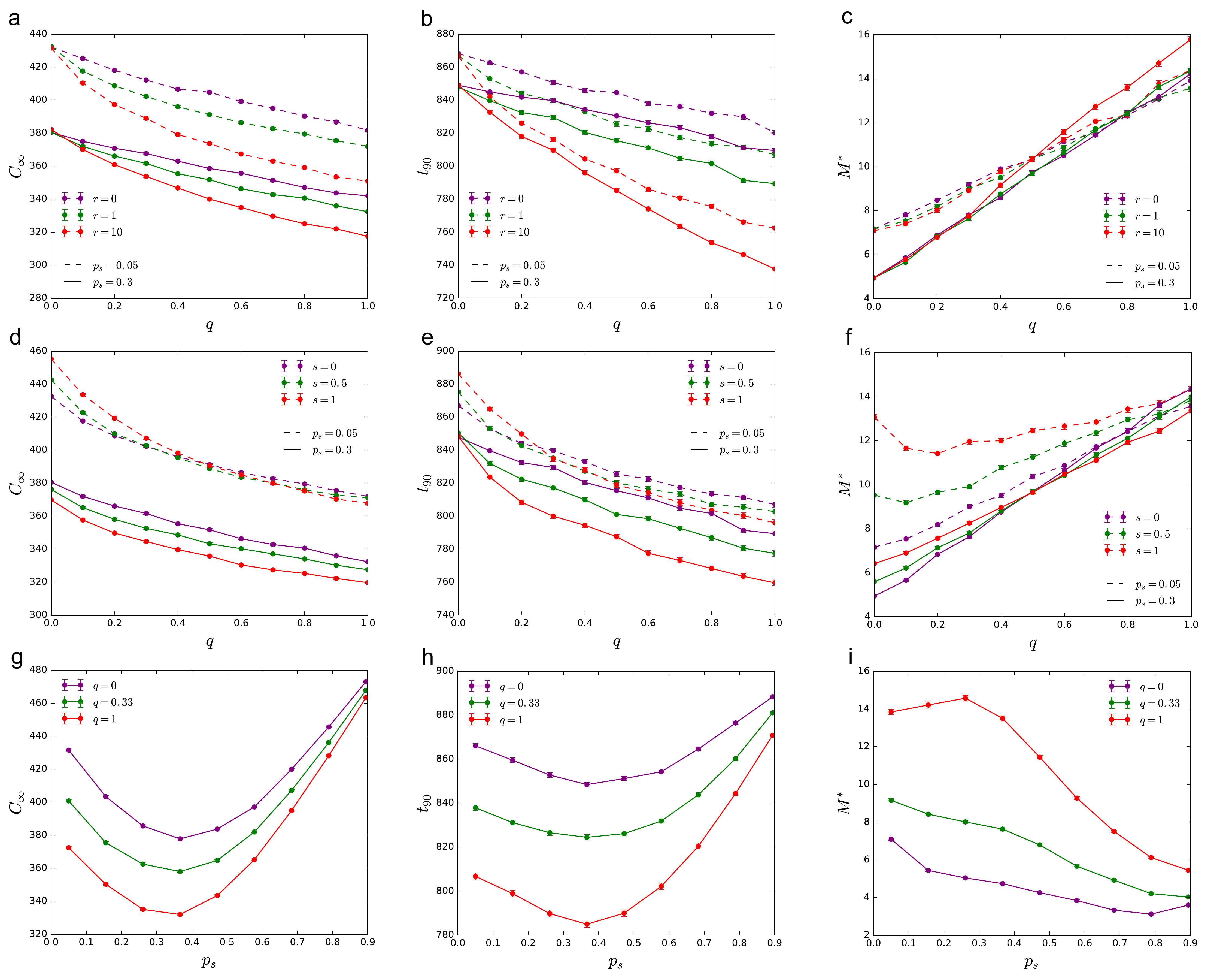}
	}	
	\caption{Effect of network structure on recovery. (a-f)~Redundancy $q$ reduces optimal cost $C_\infty$ and recovery time $t_{90}$ and increases $M^*$, the number of edges needed to be sampled to approximate the optimal solution. This effect is independent of the value of $r$, $s$, and $p_\T{s}$ (dashed and solid lines). (g-h)~Centralized and unrealistically distributed supply both increase $C_\infty$ and recovery time $t_{90}$, with a minimum at intermediately distributed supply. (i)~Increasing $p_\T s$, improves approximability. Total cost $C_\infty$ is measured in units of $(\T{total demand} \times \T{recovery timestep})$. Networks were generated using parameters $N=1000$, $q=0.33$, $r=1$, $s=0$, and $p_\T{s}=0.3$, unless otherwise indicated in the figure. Each data point is an average of $100$ independent network realizations, and the error bars indicate the standard error of the mean.}
	\label{fig:parameter_sweap}
\end{figure}

Simulations show that redundancy $q$, which controls the average degree, has the strongest effect (Figs.~\ref{fig:parameter_sweap}a-c). Increasing redundancy lowers both optimal total cost $C_\infty$ and recovery time $t_{90}$; however, it increases $M^*$, meaning that to approximate the optimal solution more edges need to be sampled. This observation is robust to the choice of other parameters. Redundancy increases possible ways to reconnect the network, allowing less costly reconstruction strategies, but this also means that more paths must be explored to pick out the optimal one. The effect of $r$ is more subtle, we find that long range shortcuts ($r=10$) further decrease $C_\infty$ and $t_{90}$; while short cycles ($r=0$) have the opposite effect. The value of $r$ has little effect on $M^*$.

The effect of line splitting depends on both the fraction of suppliers $p_\T{s}$ and the redundancy $q$ (Figs.~\ref{fig:parameter_sweap}d-f). For centralized supply ($p_\T s = 0.05$), we find that in case of low redundancy, $s$ increases cost $C_\infty$ and recovery time $t_{90}$; while in case of high redundancy,  $s$ has the opposite effect, reducing $C_\infty$ and $t_{90}$. Independent of the value of $q$, the characteristic $M^*$ is significantly increased. For distributed supply ($p_\T s = 0.3$), we find that both  $C_\infty$ and $t_{90}$ are increased by $s$ independently of the value of $q$. While the value of $M^*$ is increased by $s$ for high $q$, and decreased for low $q$.

Finally, the fraction of suppliers $p_\T s$ also strongly influences the recovery process (Figs.~\ref{fig:parameter_sweap}g-i). Total optimal cost  $C_\infty$ and recovery time $t_{90}$ are high for very centralized (low $p_\T s$) and  very distributed (high $p_\T{s}$) supply, with a minimum in between. If the demand and supply follow the same distribution, the minimum is at $p^*_\T s = 0.5$. For our choice, the demand is more heterogeneously distributed than the supply, resulting $p^*_\T s<0.5$. Increasing $p_\T s$, also allows easier approximation of the optimal solution, i.e., $M^*$ decreases with increasing $p_\T s$ (with the exception of low $q$ and $p_\T s$).

Overall, we find that high structural redundancy reduces the optimal cost and time of recovery; however, higher edge sampling $M$ is needed to benefit from this reduction. Long range shortcuts in the network further reduce the cost, without significantly increasing $M$. We also benefit from distributed supply, reducing both cost and recovery time, and depending on the level redundancy, may also improve approximability.

\section*{Discussion}

We investigated the problem of optimal cost reconstruction of critical infrastructure systems after catastrophic events. We started by analyzing realistic recovery strategies for a small-scale case study, the power grid of Shelby County, TN. We identified the penalty incurred for over- and undersupply of commodities as the main contribution to the cost, outranking operational and repair costs by orders of magnitude in the initial periods of recovery. Motivated by this observation, we introduced the recovery percolation model, a competitive percolation model that in addition to network structure also takes the demand and supply associated with each node into account. The advantage of our stylized model is that it is computationally tractable and easy to interpret compared to the complex optimization problems studied in the civil engineering literature, while adequately reproducing important features of realistic recovery processes. This allows us to identify underlying mechanisms of the recovery process. For example, we showed that component size anti-correlates with the unsatisfied demand, which suppresses the emergence of large-scale connectivity through a process analogous to explosive percolation. Such a suppression of large-scale connectivity can be in fact observed in real recovery events~\cite{Rudnick2011a}. The model also allowed us to systematically investigate the effect of typical network characteristics on the efficiency of the recovery process using synthetic power grids.

The computational complexity of identifying actionable reconstruction strategies is an open issue, especially in the case of interdependent and decentralized recovery scenarios, where systems are larger, and the optimization problem must be solved numerous times~\cite{Gonzalez2016a,Gonzalez2017a,Smith2017a}. Our stylized model is efficient, but still ignores details. Similar to td-NPD, these strategies provide scenarios that may be useful for developing recovery operator based approaches to mathematically model the dynamics of recovery and enable development of data-driven control approaches~\cite{Chapman2017a}. Further work is needed to extend our model to simultaneous recovery of multiple critical infrastructure systems explicitly taking into account interdependencies between the systems. Competitive percolation strategies in general, can provide opportunities for modeling real-world processes. For instance, in addition to this application to recovery, there is recent work of applying competitive percolation strategies to suppress the outbreak of epidemics via targeted immunization~\cite{clusella2016}. 

\begin{figure*}
\begin{minipage}{0.8\textwidth}
\begin{subequations}
\begin{flalign*}
\text{td-NDP} = \text{ minimize} &\sum_{t \in \mathcal{T} \mid t > 0}\bigg(\sum_{(i,j) \in \mathcal{E}^{\prime}}{f_{ijt}\widetilde{y}_{ijt}} + \sum_{i \in \mathcal{N}^{\prime}}{q_{it}\widetilde{w}_{it}} \bigg)&\\
& + \sum_{t \in \mathcal{T}} \bigg( {\sum_{i \in \mathcal{N}}{\boxed{(M^{+}_{it}\delta^{+}_{it} + M^{-}_{it}\delta^{-}_{it})}}} + {\sum_{(i,j) \in \mathcal{E}}{c_{ijt}x_{ijt}}}\bigg) \numberthis \label{eq:mipobj}\\
\text{subject to,}
\end{flalign*}
\begin{flalign}
 & \sum_{j:(i,j) \in \mathcal{E}}{x_{ijt}} - \sum_{j:(j,i) \in \mathcal{E}}{x_{jit}} = b_{it} - \delta^{-}_{it} + \delta^{+}_{it}, & \forall i \in \mathcal{N}, \forall t \in \mathcal{T},&
 \label{eq:mip_flow}
\end{flalign}
\begin{flalign}
 & x_{ijt} \leq u_{ijt}w_{it}, & \forall (i,j) \in \mathcal{E}, \forall t \in \mathcal{T},&
 \label{eq:mip_nodefunc}
\end{flalign}
\begin{flalign}
 & x_{ijt} \leq u_{ijt}w_{jt}, & \forall (i,j) \in \mathcal{E}, \forall t \in \mathcal{T}, &
\end{flalign}
\begin{flalign}
 & x_{ijt} \leq u_{ijt}y_{ijt}, & \forall (i,j) \in \mathcal{E}, \forall t \in \mathcal{T}, \label{eq:mip_arcfunc}&
\end{flalign}
\begin{flalign}
 & \sum_{i \in \mathcal{N}^{\prime}}\widetilde{w}_{it} + \sum_{(i,j) \in \mathcal{E}^{\prime}}\widetilde{y}_{ijt}\leq 1 & \forall t \in \mathcal{T} \mid t > 0,&
\label{eq:mip_resource}
\end{flalign}
\begin{flalign}
& w_{it}  \leq \sum_{\tilde{t}=1}^{t}{\widetilde{w}_{i\tilde{t}}},  & \forall i \in \mathcal{N}^{\prime}, \forall t \in \mathcal{T} | t > 0,&
\label{eq:mip_nodetimeconstr}
\end{flalign}
\begin{flalign}
 & y_{ijt} \leq \sum_{\tilde{t}=1}^{t}{\widetilde{y}_{ij\tilde{t}}}, & \forall (i,j) \in \mathcal{E}^{\prime}, \forall t \in \mathcal{T} | t > 0,&
 \label{eq:mip_arctimeconstr}
\end{flalign}
\begin{flalign}
& \delta^{+}_{it} \geq 0, & \forall i \in \mathcal{N}, \forall t \in \mathcal{T}, &
\label{eq:mip_variables_begin}
\end{flalign}
\begin{flalign}
& \delta^{-}_{it} \geq 0, & \forall i \in \mathcal{N}, \forall t \in \mathcal{T}, & \\
& x_{ijt} \geq 0, & \forall (i,j) \in \mathcal{E}, \forall t \in \mathcal{T}, & \\
& w_{it} \in \lbrace 0,1 \rbrace, & \forall i \in \mathcal{N}, \forall t \in \mathcal{T}, & \\
& y_{ijt} \in \lbrace 0,1 \rbrace, & \forall (i,j) \in \mathcal{E}, \forall t \in \mathcal{T}, & \\
& \widetilde{w}_{it} \in \lbrace 0,1 \rbrace, & \forall i \in \mathcal{N}^{\prime}, \forall t \in \mathcal{T}, & \\
& \widetilde{y}_{ijt} \in \lbrace 0,1 \rbrace, & \forall (i,j) \in \mathcal{E}^{\prime}, \forall t \in \mathcal{T}. &
\label{eq:mip_variables_end}
\end{flalign}
\end{subequations}
\end{minipage}
\end{figure*}

\section*{Methods}

\subsection*{Time-dependent NDP}

Here, we define our benchmark model for network recovery: the time-dependent network design problem (td-NDP). Our version follows the more general formulation developed by Gonzalez et al.~\cite{Gonzalez2016a,Gonzalez2016b,Gonzalez2017a}. The td-NDP takes a graph $\mathcal{G}=(\mathcal{N},\mathcal{E})$, where $\mathcal{N}$ is a set of nodes, and $\mathcal{E}$ is the set of edges connecting nodes. At the beginning of the recovery process the td-NDP uses the destroyed graph, $\mathcal{G}^{\prime}=(\mathcal{N}^{\prime},\mathcal{E}^{\prime})$, where $\mathcal{N}^{\prime}$ and $\mathcal{E}^{\prime}$ represents the nodes and edges that are not functioning, respectively. The objective function (cf.~Equation~\eqref{eq:mipobj}) minimizes the total reconstruction cost over a given time domain $\mathcal{T}$ with $t\in \mathcal{T}$, which includes the cost to repair nodes, $q_{it}$, cost to repair edges, $f_{ijt}$, cost of flow on each edge, $c_{ijt}$, and oversupply and undersupply penalties for each node, $M^{+}_{it}$ and $M^{-}_{it}$. These costs usually depend on multiple factors, such as the level of damage, the type and size of the components to be restored, their geographical accessibility, the amount of labor and resources required, and the social vulnerability of the affected areas, among others~\cite{Gonzalez2016a,hazus,Hutcheon2013a}. To keep track of demand satisfaction, each node also has a supply capacity (demand if negative), $b_{it}$. In the most general formalization of the problem, node supply $b_{it}$ can depend on time $t$, but in this paper we only consider constant values. The variables $\delta^{+}_{it}$ ($\delta^{-}_{it}$) account for oversupply (or undersupply) of node $i$. We refer to the sum of the absolute values of oversupply and undersupply ($\left| \delta^{+}_{it} \right|+\left| \delta^{-}_{it} \right|$) as the commodity deficit of node $i$. The td-NDP includes as decision variables the amount of flow on each edge, $x_{ijt}$, whether or not a node $i$ [edge $(i,j)$] is chosen to be recovered at timestep $t$, $\tilde{w}_{it}$ ($\tilde{y}_{ijt}$), and whether or not a node $i$ [edge $(i,j)$] is functional at timestep $t$, $w_{it}$ ($y_{ijt}$). Constraints~\ref{eq:mip_flow}-~\ref{eq:mip_variables_end} are imposed to ensure that conservation of flow properties are held and that only recovered and functional nodes can produce or consume flow.

The td-NDP formulation is a mixed integer program, which has been shown to be, in general, NP-hard (and becomes exponentially harder as $\mathcal{T}$ and $\left| \mathcal{G}^{\prime} \right|$ grows). The number of variables and constraints also become larger as the input graph becomes larger. For many reasonable size problems, computing a global optimal (i.e., where $\mathcal{T}$ contains the entire time horizon for recovery) is intractable. Therefore, heuristics are used to restrict the size of $\mathcal T$ by dividing the total recovery time into smaller windows, and finding the locally optimal solutions within these windows~\cite{Gonzalez2016a}. It has been shown that such heuristic finds solutions very close to the optimal; however, the computational complexity is still relatively high as a result of the underlying mixed-integer program.

\subsection*{Supply and demand distribution}
For our computational experiments, we generate our demand distribution by following the load distribution of the European power grid~\cite{Hutcheon2013a}. This dataset was chosen due to its large system size ($N=1463,E=2199$) and its high resolution. Our goal is not to identify the true analytic form of the load distribution, but to generate statistically similar samples through bootstrapping. We found that an exponentiated Weibull distribution of the form $f(x,a,c)=ac(1-\exp{(-x^c)}^{(a-1)}\exp{(-x^c)}x^{(c-1)})$, where $a=3.59$ and $c=0.8$ well approximates the features of the demand distribution. Suppliers' capacities are uniformly distributed to balance the total demand. We also get our ratio of suppliers ($0.3$) to consumers ($0.7$) from this dataset.

\section*{Acknowledgements}
We gratefully acknowledge support from the U.S. Army Research Laboratory and the U.S. Army Research Office under MURI award number W911NF-13-1-0340, and from DARPA award W911NF-17-1-0077.

\section*{Author contributions statement}

A.S. was responsible for experimental design and analysis; M.P. was responsible for problem scoping and providing theoretical insights; M.R. provided guidance for power grid analysis and network formation; A.G. and L.D.O. contributed recovery and optimization algorithms and background; R.D. provided percolation and network science guidance and expertise. All authors contributed to the transcription of this article.

\section*{Competing interests}
The author(s) declare no competing interests.

\end{document}